\documentclass[twocolumn,showpacs]{revtex4}
\usepackage{graphicx}

\begin{document}
\title{Angular momentum conservation for dynamical black holes}
\author{Sean A. Hayward}
\affiliation{Center for Astrophysics, Shanghai Normal University, 100 Guilin
Road, Shanghai 200234, China}
\date{Revised 20th October 2006}

\begin{abstract}
Angular momentum can be defined by rearranging the Komar surface integral in
terms of a twist form, encoding the twisting around of space-time due to a
rotating mass, and an axial vector. If the axial vector is a coordinate vector
and has vanishing transverse divergence, it can be uniquely specified under
certain generic conditions. Along a trapping horizon, a conservation law
expresses the rate of change of angular momentum of a general black hole in
terms of angular momentum densities of matter and gravitational radiation.
This identifies the transverse-normal block of an effective
gravitational-radiation energy tensor, whose normal-normal block was recently
identified in a corresponding energy conservation law. Angular momentum and
energy are dual respectively to the axial vector and a previously identified
vector, the conservation equations taking the same form. Including charge
conservation, the three conserved quantities yield definitions of an effective
energy, electric potential, angular velocity and surface gravity, satisfying a
dynamical version of the so-called first law of black-hole mechanics. A
corresponding zeroth law holds for null trapping horizons, resolving an
ambiguity in taking the null limit.
\end{abstract}
 \maketitle

\section{Introduction}
The theory of black holes appears finally to be reaching a stage of maturity
in which it can be applied in the most interesting, distorted, dynamic
situations, with appropriate definitions of relevant physical quantities. This
article mainly concerns angular momentum and its conservation, which is the
last major piece of what seems to be an essentially complete new paradigm for
black holes. It therefore seems timely to review below the key ideas and
results of what might be called the heroic, classical and modern eras.

The first solution which would nowadays be called a black hole was discovered
by Schwarzschild \cite{Sch} within a few weeks of the final formulation of the
field equations of General Relativity by Einstein \cite{Ein}, as the external
gravitational field of a point with mass $M$. Charge $Q$ was soon added by
Reissner \cite{Rei} and Nordstr\"om \cite{Nor}, but even the Schwarzschild
solution was not properly understood for decades. Schwarzschild described the
mass point as located at what is now understood as the horizon, despite its
non-zero area $A$. Einstein \& Rosen \cite{ER} realized that the spatial
geometry had a wormhole structure, extending through a minimal surface.
Oppenheimer \& Snyder \cite{OS} constructed a model of stellar collapse in
which the star collapses through the horizon. Finally Kruskal \cite{Kru}, as
reported in a paper actually written by Wheeler \cite{WF}, described the
entire space-time, whence it became clear that there was a trapped region
inside the horizon, from which nothing could escape to the exterior. Angular
momentum $J$ was added by Kerr \cite{Ker} and, including $Q$, by Newman et
al.\ \cite{New}. Uniqueness theorems identify these as the only black holes
which are stationary, asymptotically flat, electrovac solutions.

Wheeler \cite{Whe} is credited with coining the term ``black hole'' and
Penrose \cite{Pen0} with defining event horizons, which became the accepted
definition of black holes. Hawking \cite{Haw0} showed that the area of an
event horizon was non-decreasing, $A'\ge0$. The result became known as a
``second law'', due to inaccurate analogies with the laws of thermodynamics
and the results summarized by Bardeen et al.\ \cite{BCH}: a ``zeroth law''
that surface gravity $\kappa$ is constant on a stationary black hole; a
``first law''
\begin{equation}
\delta E=\kappa\delta A/8\pi+\Omega\delta J+\Phi\delta Q\label{g0}
\end{equation}
for perturbations of stationary black holes, where $\Omega$ is the angular
speed and $\Phi$ the electric potential of the horizon, and the ADM energy $E$
measures the total mass of the space-time; and a ``third law'' that
$\kappa\not\to0$ by positive-energy perturbations of stationary black holes.
This summarizes the classical theory of black holes as described in textbooks
\cite{HE,MTW,Wal}.

The last results above are perhaps best described as black-hole statics, being
properties of stationary black holes, specifically of Killing horizons, rather
than of general event horizons. While adequate in some astrophysical
situations, this classical theory is inapplicable to general dynamical
processes, for instance black-hole formation, rapid evolution and binary
mergers. A theory of black-hole dynamics is needed, with generalizations of
all the above-mentioned quantities. Event horizons are not an appropriate
platform, since they cannot be located by mortals, let alone admit physical
measurements. A more practical way to locate a black hole is by a {\em
marginal surface}, an extremal surface of a null hypersurface, where light
rays are momentarily caught by the gravitational field. Marginal surfaces are
used extensively in numerical simulations, where they have historically been
called apparent horizons, though the textbook definition of the latter is
different \cite{HE,MTW,Wal}. Here a hypersurface foliated by marginal surfaces
will be called a {\em trapping horizon}.

A systematic treatment of trapping horizons \cite{bhd} distinguished four
sub-classes, called future or past, outer or inner trapping horizons, with the
future outer type proposed as the practical location of a black hole. Such a
horizon was shown to have several expected properties of a black hole,
assuming the Einstein equation and positive-energy conditions: the horizon is
achronal, being null in a special case of quasi-stationarity, but otherwise
being spatial; the area form ${*}1$ of the marginal surfaces is constant in
the null case and increasing in the spatial case; and the marginal surfaces
have spherical topology, if compact. The area law implies that the area
$A=\oint_S{*}1$ of the marginal surfaces $S$ is non-decreasing,
\begin{equation}
L_\xi A\ge0\label{a0}
\end{equation}
where $L$ denotes the Lie derivative and $\xi$ the generating vector of the
marginal surfaces. So a black hole grows if something falls into it, otherwise
staying the same size.

Comprehensive treatments were subsequently given for spherical symmetry
\cite{sph,1st,in}, cylindrical symmetry \cite{cyl} and a quasi-spherical
approximation \cite{qs,SH,gwbh,gwe}. In each case, definitions were found for
all the non-zero physical quantities mentioned above, providing prototypes of
all except $J$ and $\Omega$. In addition, an effective energy tensor $\Theta$
for gravitational radiation was found, entering equations additively to the
matter energy tensor $T$. The Einstein equations were decomposed into forms
with manifest physical meaning, such as a quasi-Newtonian gravitational law, a
wave equation for the gravitational radiation, and an energy conservation law
which can be written in the form
\begin{equation}
L_\xi
M=\oint_S{*}(T_{\alpha\beta}+\Theta_{\alpha\beta})k^\alpha\tau^\beta\label{ec0}
\end{equation}
where $\tau$ in the normal dual of $\xi$ and $k$ is a certain vector, playing
the role of a Killing vector, which is null on the horizon. Such equations
actually hold not just on a trapping horizon, but anywhere in the space-time,
energy conservation reducing at null infinity to the Bondi energy equation.

Contemporaneously, Ashtekar et al.\ and others
\cite{ABF1,ABF2,AFK,ABD,ABL1,Boo,ABL2,DKSS,GJ1} developed a theory of null
trapping horizons with various additional conditions, under the names
non-rotating isolated horizons, non-expanding horizons, weakly isolated
horizons, rigidly rotating horizons and (strongly) isolated horizons. Each is
intended to capture the idea that the black hole is quasi-stationary in some
sense. They gave definitions of all the relevant physical quantities and
derived a generalized version of the so-called first law.

Subsequently, Ashtekar \& Krishnan \cite{AK1,AK2,AK3} studied future spatial
trapping horizons under the name dynamical horizons, giving classes of
definitions of energy and angular momentum, deriving corresponding flux
equations and obtaining a version of the so-called first law for $Q=0$.
However, the ``3+1'' formalism used to describe spatial trapping horizons
breaks down when the horizon becomes null, so that the isolated-horizon and
dynamical-horizon frameworks were essentially distinct. Some connections were
drawn, particularly for slowly evolving horizons by Booth \& Fairhurst
\cite{BF1,BF2,BF3}. Recently, Andersson et al.\ \cite{AMS} showed that a
stable trapping horizon is, on any one marginal surface, either spatial or
null everywhere, so that transitions between the two types happen
simultaneously on a marginal surface. They and Ashtekar \& Galloway \cite{AG}
also obtained some existence and uniqueness results for trapping horizons.

A unified framework for any trapping horizon is provided by a dual-null
formalism \cite{dn,dne}, which was used throughout the earlier studies
\cite{bhd,sph,1st,in,cyl,qs,SH,gwbh,gwe}. The energy flux equation was then
cast in a surface-integral form where the null limit could be taken
\cite{bhd2,bhd3}. Moreover, it was cast in the form of a conservation law
(\ref{ec0}), by identifying an effective energy tensor $\Theta$. The mass $M$,
which might take any value on a given $S$ by choice of scaling of $k$, was
chosen to be the irreducible mass or Hawking mass \cite{Haw} for consistency
with the earlier studies. This corresponds to the simplest general definition
of $k$, such that it becomes a unit vector for round spheres near infinity.

The main task here is to make similar refinements for angular momentum, as
briefly described in shorter articles \cite{j15,bhd4}. In particular, one
desires not just a flux equation but a conservation law of the form
\begin{equation}
L_\xi
J=-\oint_S{*}(T_{\alpha\beta}+\Theta_{\alpha\beta})\psi^\alpha\tau^\beta.\label{amc0}
\end{equation}
Here $\psi$ should be an axial vector in some sense, playing the role of an
axial Killing vector, with $J$ being the angular momentum about that axis. It
turns out that natural restrictions on $\psi$ allow it to be uniquely
specified under certain generic conditions. The angular momentum, initially a
functional $J[\psi]$, is obtained directly from the Komar integral \cite{Kom}
in terms of a 1-form $\omega$ known as the twist \cite{dne}, which reduces to
the 1-form used for dynamical horizons \cite{AK1,AK2,AK3}. It encodes the
rotational frame-dragging predicted in the Lense-Thirring effect, thereby
giving a precise meaning to the twisting around of space-time due to a
rotating mass.

The null limit is more subtle for angular momentum than for energy, where the
irreducible mass is uniquely defined for a null trapping horizon, energy
conservation (\ref{ec0}) reducing correctly to $L_\xi M=0$. The dual-null
foliation becomes non-unique for a null hypersurface, with $\omega$ becoming
non-unique. This is reflected in the fact that different 1-forms were used for
isolated horizons \cite{ABF1,ABF2,AFK,ABD,ABL1,Boo,ABL2,DKSS,GJ1} and
dynamical horizons \cite{AK1,AK2,AK3}. Neither 1-form is necessarily preserved
along a null trapping horizon, but a certain linear combination is so
preserved. However, all three 1-forms coincide if the non-uniqueness of the
dual-null foliation for a null hypersurface is fixed in a certain way. Then
$J[\psi]$ becomes unique for a null trapping horizon and conservation
(\ref{amc0}) reduces as desired to $L_\xi J=0$. Thus a consistent treatment of
angular momentum naturally resolves the issue of the degeneracy of the null
limit. It follows that a black hole cannot change its angular momentum without
increasing its area.

The article is organized as follows. Section II summarizes the underlying
geometry. Section III reviews trapping horizons and conservation of energy.
Section IV derives angular momentum from the Komar integral and shows how
restrictions on the axial vector can be used to construct a unique definition.
Section V derives and discusses the conservation law. Section VI includes
charge conservation and discusses local versus quasi-local conservation.
Section VII describes the state space, defining the remaining physical
quantities and deriving a dynamical version of the so-called first law.
Section VIII considers null trapping horizons, deriving a zeroth law.
Appendices concern (A) weak fields, (B) normal fundamental forms and (C) a
Kerr example. The above discussion serves as a summary.

\section{Geometry}
\begin{figure}
\includegraphics[height=15mm]{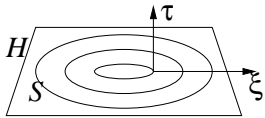}
\caption{A non-null hypersurface $H$ foliated by spatial surfaces $S$, with
generating vector $\xi$ and its normal dual $\tau=\xi^*$. If $H$ becomes null,
$\xi$ and $\tau$ coincide.} \label{normal}
\end{figure}

General Relativity will be assumed, with space-time metric $g$. The
geometrical object of interest is a one-parameter family $\{S\}$ of spatial
surfaces $S$, locally generating a foliated hypersurface $H$. Labelling the
surfaces by a coordinate $x$, they are generated by a vector
$\xi=\partial/\partial x$, which can be taken to be normal to the surfaces,
$\bot\xi=0$, where $\bot$ denotes projection onto $S$. A Hodge duality
operation on normal vectors $\eta$, $\bot\eta=0$, yields a dual normal vector
$\eta^*$ satisfying
\begin{equation}
\bot\eta^*=0,\quad g(\eta^*,\eta)=0,\quad
g(\eta^*,\eta^*)=-g(\eta,\eta).\label{nd}
\end{equation}
In particular,
\begin{equation}
\tau=\xi^*
\end{equation}
is normal to $H$, with the same scaling (Fig.\ref{normal}). The coordinate
freedom here is just $x\mapsto\tilde x(x)$ and choice of transverse
coordinates on $S$, under which all the key formulas will be invariant. The
generating vector $\xi$ may have any causal character, at each point. For
instance, a future outer trapping horizon is spatial while growing, becomes
null when quasi-stationary, and would become temporal if shrinking during
evaporation \cite{bhd,bhd2,bhd3}.

A dual-null formalism \cite{dn,dne} describes two families of null
hypersurfaces $\Sigma_\pm$, intersecting in a two-parameter family of spatial
surfaces, including the desired one-parameter family. Some merits of the
dual-null approach, apart from comparative ease of calculation, are that it is
adapted both to marginal surfaces, defined as extremal surfaces of null
hypersurfaces, and to radiation propagation, which makes it easier to identify
terms arising due to gravitational radiation \cite{bhd2,bhd3,bhd4}. Relevant
aspects of the formalism are summarized as follows.

Labelling $\Sigma_\mp$ by coordinates $x^\pm$ which increase to the future,
one may take transverse coordinates $x^a$ on $S$, which for a sphere would
normally be angular coordinates $x^a=(\vartheta,\varphi)$. Writing space-time
coordinates $x^\alpha=(x^+,x^-,x^a)$ indicates how one may use Greek letters
$(\alpha,\beta,\ldots)$ for space-time indices and corresponding Latin letters
$(a,b,\ldots)$ for transverse indices. The coordinate basis vectors are
$\partial_\alpha=\partial/\partial x^\alpha$ and the dual 1-forms are
$dx^\alpha$, satisfying $\partial_\beta(dx^\alpha)=\delta^\alpha_\beta$.
Coordinate vectors commute, $[\partial_\alpha,\partial_\beta]=0$, where the
brackets denote the Lie bracket or commutator. Two coordinate vectors have a
special role, the evolution vectors $\partial_\pm=\partial/\partial x^\pm$
which generate the dynamics, spanning an integrable evolution space. The
corresponding normal 1-forms $dx^\pm$ are null by assumption:
\begin{equation}
g^{-1}(dx^\pm,dx^\pm)=0.
\end{equation}
The relative normalization of the null normals may be encoded in a function
$f$ defined by
\begin{equation}
e^f=-g^{-1}(dx^+,dx^-)\label{ff}
\end{equation}
where the metric sign convention is that spatial metrics are positive
definite. The transverse metric, or the induced metric on $S$, is found to be
\begin{equation}
h=g+2e^{-f}dx^+\otimes dx^-\label{tm}
\end{equation}
where $\otimes$ denotes the symmetric tensor product. There are two shift
vectors
\begin{equation}
s_\pm=\bot\partial_\pm
\end{equation}
where $\bot$ is generalized to indicate projection by $h$. The null normal
vectors
\begin{equation}
l_\pm=\partial_\pm-s_\pm=-e^{-f}g^{-1}(dx^\mp)\label{nn}
\end{equation}
are future-null and satisfy
\begin{eqnarray}
&&g(l_\pm,l_\pm)=0,\quad g(l_+,l_-)=-e^{-f},\nonumber\\
&&l_\pm(dx^\pm)=1,\quad l_\pm(dx^\mp)=0,\quad\bot l_\pm=0.\label{ls}
\end{eqnarray}
The metric takes the form
\begin{eqnarray}
g&=&h_{ab}(dx^a+s_+^adx^++s_-^adx^-)\otimes\nonumber\\&&(dx^b+s_+^bdx^++s_-^bdx^-)
-2e^{-f}dx^+\otimes dx^-.
\end{eqnarray}
Then $(h,f,s_\pm)$ are configuration
fields and the independent momentum fields are found to be linear combinations
of the following transverse tensors:
\begin{eqnarray}
\theta_\pm&=&{*}L_\pm{*}1\label{ex}\\ \sigma_\pm&=&\bot L_\pm h-\theta_\pm h\label{sh}\\
\nu_\pm&=&L_\pm f\label{in}\\
\omega&=&{\textstyle\frac12}e^fh([l_-,l_+])\label{tw}
\end{eqnarray}
where ${*}$ is the Hodge operator of $h$ and $L_\pm$ is shorthand for the Lie
derivative along $l_\pm$. Then the functions $\theta_\pm$ are the null
expansions, the traceless bilinear forms $\sigma_\pm$ are the null shears, the
1-form $\omega$ is the twist, measuring the lack of integrability of the
normal space, and the functions $\nu_\pm$ are the inaffinities, measuring the
failure of the null normals to be affine. The fields
$(\theta_\pm,\sigma_\pm,\nu_\pm,\omega)$ encode the extrinsic curvature of the
dual-null foliation. These extrinsic fields are unique up to interchange
$\pm\mapsto\mp$ and diffeomorphisms $x^\pm\mapsto\tilde x^\pm(x^\pm)$ which
relabel the null hypersurfaces. Further description of the geometry was given
recently \cite{bhd3}.

As described, the dual-null formalism is manifestly covariant on $S$, with
transverse indices not explicitly denoted, while $\pm$ indices indicate the
chosen normal basis \cite{dne,bhd3}. Conversely, one can use a formalism which
is manifestly covariant on the normal space, with transverse but not normal
indices explicitly denoted \cite{BHMS}. Both types of formalism can seem
obscure to the uninitiated, so indices will be explicitly denoted in longer
formulas in this article, nevertheless being omitted where the meaning should
be clear. Capital Latin letters $(A,B,\ldots)$ will be used for normal
indices, when not denoted by $\pm$ in the dual-null basis. Then the
configuration fields are $(h_{ab},f,s_A{}^b)$, the momentum fields are
$(\theta_A,\sigma_{Abc},\nu_A,\omega_a)$ and the derivative operators are
$(\bot L_A,D_a)$, where $D$ is the covariant derivative operator of $h$.

Since the normal space is not integrable unless $\omega=0$, it generally does
not admit a coordinate basis. However, one may still take $dx^\pm$ as basis
1-forms, in which case the dual basis vectors are $l_\pm$, as follows from
(\ref{ls}), implying $l_A(dx^B)=\delta_A^B$. In this basis, the normal metric
\begin{equation}
\gamma=g-h
\end{equation}
has components which follow from (\ref{tm}) as
\begin{equation}
\gamma_{AB}=-e^{-f}(dx^+_Adx^-_B+dx^-_Adx^+_B)
\end{equation}
and its inverse has components
\begin{equation}
\gamma^{AB}=-e^f(l_+^Al_-^B+l_-^Al_+^B)
\end{equation}
which can be used to lower and raise normal indices. Also useful is the
binormal
\begin{equation}
\epsilon_{AB}=e^{-f}(dx^+_Adx^-_B-dx^-_Adx^+_B) \label{bnd}
\end{equation}
or its inverse
\begin{equation}
\epsilon^{AB}=e^f(l_-^Al_+^B-l_+^Al_-^B). \label{bnu}
\end{equation}
The mixed form
\begin{equation}
\epsilon^A{}_B=l_+^Adx^+_B-l_-^Adx^-_B \label{bnm}
\end{equation}
has components $\epsilon^\pm{}_\pm=\pm1$, $\epsilon^\pm{}_\mp=0$, so can be
used to express the duality operation (\ref{nd}) on normal vectors, extended
to the dual-null foliation, as
\begin{equation}
(\eta^*)^A=\epsilon^A{}_B\eta^B. \label{nd2}
\end{equation}

The dual-null Hamilton equations and integrability conditions for vacuum
Einstein gravity were derived previously \cite{dne}, with matter terms added
subsequently \cite{gwbh}. The components of the field equations which are
relevant to angular momentum turn out to be the twisting equations
\begin{eqnarray}
\bot L_\pm\omega_a&=&-\theta_\pm\omega_a\pm\textstyle{\frac12}D_a\nu_\pm
\mp\textstyle{\frac12}D_a\theta_\pm\mp\textstyle{\frac12}\theta_\pm
D_af \nonumber\\
&&\qquad\qquad{}\pm\textstyle{\frac12}h^{cd}D_d\sigma_{\pm ac} \mp8\pi
T_{a\pm}\label{twisting}
\end{eqnarray}
where $T_{a\pm}=h_a^\gamma T_{\gamma\beta}l_\pm^\beta$ is the
transverse-normal projection of the energy tensor $T$, and units are such that
Newton's gravitational constant is unity. The corresponding all-index version
can be written using the binormal as
\begin{eqnarray}
\bot
L_B\omega_a&=&-\theta_B\omega_a+\textstyle{\frac12}\epsilon^E{}_B(D_a\nu_E
-D_a\theta_E-\theta_ED_af \nonumber\\
&&\qquad\qquad\quad{}+h^{cd}D_d\sigma_{Eac}-16\pi T_{aE}).
\end{eqnarray}

\section{Trapping horizons and conservation of energy}
Returning to a general foliated hypersurface $H$, a normal vector $\eta$ has
components $\eta^\pm$ along $l_\pm$, so that $\eta=\eta^+l_++\eta^-l_-$, and
its normal dual is $\eta^*=\eta^+l_+-\eta^-l_-$. In particular, the generating
vector is
\begin{equation}
\xi=\xi^+l_++\xi^-l_-\label{xi}
\end{equation}
and its dual is
\begin{equation}
\tau=\xi^+l_+-\xi^-l_-.\label{tau}
\end{equation}
Since the horizon is given parametrically by functions $x^\pm(x)$, the
components $\xi^\pm=\partial x^\pm/\partial x$ are independent of transverse
coordinates:
\begin{equation}
D\xi^\pm=0.\label{dxi}
\end{equation}
It is also useful to introduce the expansion
\begin{equation}
\theta_\eta=L_\eta\log({*}1)=\theta_A\eta^A
\end{equation}
along a normal vector $\eta$, particularly the expansion $\theta_\xi$ along
the generating vector.

A trapping horizon \cite{bhd,1st,bhd2,bhd3} is a hypersurface $H$ foliated by
marginal surfaces, where $S$ is marginal if one of the null expansions,
$\theta_+$ or $\theta_-$, vanishes everywhere on $S$. Then $S$ is an extremal
surface of the null hypersurface $\Sigma_+$ or $\Sigma_-$.

The recently derived energy conservation law \cite{bhd2,bhd3} will be stated
here for later comparison, modifying some notation. Assuming compact $S$
henceforth, the transverse surfaces have area
\begin{equation}
A=\oint_S{*}1
\end{equation}
and the area radius
\begin{equation}
R=\sqrt{A/4\pi}
\end{equation}
is often more convenient. The Hawking mass \cite{Haw}
\begin{equation}
M=\frac
R2\left(1-\frac1{16\pi}\oint_S{*}\gamma^{AB}\theta_A\theta_B\right)\label{hm}
\end{equation}
can be used as a measure of the active gravitational mass on a transverse
surface. Assuming the null energy condition, $M$ is the irreducible mass $R/2$
of a future outer trapping horizon, $L_\xi M\ge0$, by the area law (\ref{a0}).
On a stationary black-hole horizon, $M$ reduces to the usual definition of
irreducible mass for a Kerr-Newman black hole, namely the mass which must
remain even if rotational or electrical energy is extracted. It is generally
not the ADM energy, but an effective energy $E$ is defined in \S VII which
does recover the ADM energy in this case. Equality on a trapping horizon will
be denoted by $\cong$, e.g.\ $R\cong2M$.

Mass or energy has a certain duality with time, e.g.\ there is a standard
formula for energy if a stationary Killing vector exists. For a general
compact surface, the simplest definition of such a vector which applies
correctly for a Schwarzschild black hole is \cite{bhd2,bhd3}
\begin{equation}
k=(g^{-1}(dR))^*\label{time}
\end{equation}
or $k^A=\epsilon^{AB}L_BR$. This vector actually was found to be the
appropriate dual of $M$, in the sense of conservation laws for trapping
horizons \cite{bhd2,bhd3} and for uniformly expanding flows \cite{gr,BHMS}. In
either case, the energy conservation law can be written as
\begin{equation}
L_\xi M\cong\oint_S{*}(T_{AB}+\Theta_{AB})k^A\tau^B \label{ec}
\end{equation}
where $\Theta$ is an effective energy tensor for gravitational radiation. This
determines only the normal-normal components of $\Theta$, as
\begin{eqnarray}
\Theta_{\pm\pm}&=&||\sigma_\pm||^2/32\pi\label{Theta0}\\
\Theta_{\pm\mp}&=&e^{-f}|\omega\mp{\textstyle\frac12}Df|^2/8\pi\label{Theta1}
\end{eqnarray}
where $|\zeta|^2=h^{ab}\zeta_a\zeta_b$ and
$||\sigma||^2=h^{ac}h^{bd}\sigma_{ab}\sigma_{cd}$. Further discussion is
referred to \cite{bhd2,bhd3,gr,BHMS}.

\section{Angular momentum}
\begin{figure}
\includegraphics[height=3cm]{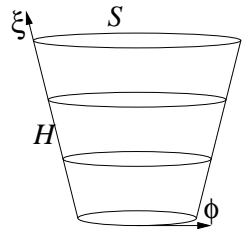}
\caption{A transverse vector $\psi$.} \label{transverse}
\end{figure}

The standard definition of angular momentum for an axial Killing vector $\psi$
and at spatial infinity is the Komar integral \cite{Kom}
\begin{equation}
J[\psi]=-\frac1{16\pi}\oint_S{*}\epsilon_{\alpha\beta}\nabla^\alpha\psi^\beta.
\label{komar}
\end{equation}
Now consider $\psi$ to be a general transverse vector, $\bot\psi=\psi$
(Fig.\ref{transverse}). Since $\epsilon_{\alpha\beta}\psi^\beta=0$, the Komar
integral can be rewritten via (\ref{bnu}) as
\begin{equation}
J[\psi]=\frac1{8\pi}\oint_S{*}\psi^a\omega_a \label{am}
\end{equation}
where $\omega$ is the twist (\ref{tw}). Since the twist encodes the
non-integrability of the normal space, it provides a geometrical measure of
rotational frame-dragging. It is an invariant of a dual-null foliation and
therefore of a non-null foliated hypersurface $H$, so the twist expression for
$J[\psi]$ is also an invariant. Appendix A shows that $J[\psi]$ recovers the
standard definition of angular momentum for a weak-field metric \cite{MTW},
with the twist being directly related to the precessional angular velocity of
a gyroscope due to the Lense-Thirring effect. Thus the twist does indeed
encode the twisting around of space-time caused by a rotating mass.

There are several definitions of angular momentum which are similar surface
integrals of an axial vector contracted with a 1-form \cite{ABD,AK1,BY}, the
situation being clarified by Gourgoulhon \cite{Gou} and in Appendix B.
Ashtekar \& Krishnan \cite{AK1} gave a definition for a dynamical horizon
which involves a 1-form coinciding with $\omega$. Brown \& York \cite{BY} gave
a definition which was stated only for an axial Killing vector $\psi$,
involving a 1-form which is generally inequivalent to $\omega$, but can be
made to coincide if adapted to a trapping horizon. Ashtekar et al.\
\cite{ABD,ABL1} gave a definition for a type II (rigidly rotating) isolated
horizon, using a 1-form which is generally inequivalent to $\omega$. However,
it can be made to coincide with $\omega$ if the dual-null foliation is fixed
in a natural way, as described in the penultimate section.

The above properties suggest (\ref{am}) as a general quasi-local definition of
angular momentum. However, if the transverse vector $\psi$ does not have
properties expected of an axial vector, the physical interpretation as angular
momentum is questionable. For instance, it would be natural to expect an axial
vector to have integral curves which form a smooth foliation of circles, apart
from two poles, assuming spherical topology for $S$. In the following, two
conditions on $\psi$ with various motivations will be considered, which, taken
together, determine $\psi$ uniquely in a certain generic situation. These
conditions then yield a conservation law with the desired form (\ref{amc0}),
as described in the next section.

Ashtekar \& Krishnan \cite{AK2} proposed that $\psi$ has vanishing transverse
divergence:
\begin{equation}
D_a\psi^a\cong0. \label{div}
\end{equation}
This condition holds if $\psi$ is an axial Killing vector, and can be
understood as a weaker condition, equivalent to $\psi$ generating a symmetry
of the area form rather than of the whole metric, since
$L_\psi({*}1)={*}D_a\psi^a$. Alternatively, assuming that the integral curves
of $\psi$ are closed, it can always be satisfied by choice of scaling of
$\psi$, as discussed by Booth \& Fairhurst \cite{BF2}. The original motivation
was that the different 1-forms used for dynamical and isolated horizons,
denoted here by $\omega$ and $\omega+\frac12Df$, will then give the same
angular momentum $J[\psi]$, by the Gauss divergence theorem.

\begin{figure}
\includegraphics[height=25mm]{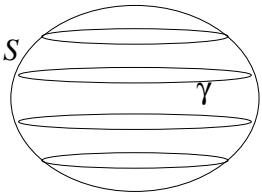}
\caption{Curves $\gamma$ of constant expansion $\theta_\xi$.} \label{axial}
\end{figure}

Spherical topology will be assumed henceforth for $S$, which follows from the
topology law \cite{bhd} for outer trapping horizons, assuming the dominant
energy condition. If there exist angular coordinates $(\vartheta,\varphi)$ on
$S$ with $\psi=\partial/\partial\varphi$, completing coordinates
$(x,\vartheta,\varphi)$ on $H$, then since coordinate vectors commute,
\begin{equation}
L_\xi\psi\cong0. \label{lie}
\end{equation}
This condition was proposed as a natural way to propagate $\psi$ along $H$ by
Gourgoulhon \cite{Gou}. Now there is a commutator identity \cite{dne}
\begin{equation}
L_\xi(D_a\psi^a)-D_a(L_\xi\psi)^a=\psi^aD_a\theta_\xi. \label{com}
\end{equation}
Therefore assuming both conditions (\ref{div})--(\ref{lie}) forces
\begin{equation}
\psi^aD_a\theta_\xi\cong0. \label{con}
\end{equation}
This is automatically satisfied if $D\theta_\xi\cong0$, as in spherical
symmetry or along a null trapping horizon. However, generically one expects
$D\theta_\xi\not\cong0$ almost everywhere. It must vanish somewhere on a
sphere, by the hairy ball theorem, but the simplest generic situation is that
there are curves $\gamma$ of constant $\theta_\xi$ which form a smooth
foliation of circles, covering the surface except for two poles
(Fig.\ref{axial}). Assuming so, since $\psi$ is tangent to $\gamma$, one can
find a unique $\psi$, up to sign, in terms of the unit tangent vector
$\hat\psi$ and arc length $ds$ along $\gamma$:
\begin{equation}
\psi\cong\hat\psi\oint_\gamma ds/2\pi \label{un}
\end{equation}
where the scaling ensures that the axial coordinate $\varphi$ is identified at
0 and $2\pi$. Then the angular momentum becomes unique up to sign,
$J[\psi]=J$, the sign being naturally fixed by $J\ge0$ and continuity of
$\psi$, corresponding to a choice of orientation.

For an axisymmetric space-time with axial Killing vector $\psi$, (\ref{div})
holds, while (\ref{lie}) holds if $\xi$ respects the symmetry, $L_\psi\xi=0$,
so the above construction, if unique as assumed, yields the correct axial
vector. In particular, the construction does work for a Kerr space-time, as
described in Appendix C.

To summarize this section, the definition (\ref{am}) of angular momentum can
be made generically unique if the axial vector is a coordinate vector,
(\ref{lie}), and generates a symmetry of the area form, (\ref{div}). The
construction can be applied in any situation where $D\theta_\xi\not\cong0$
almost everywhere, though the physical interpretation as angular momentum
seems to be safest in the case of two poles, which locate the axis of
rotation. Then $J$ is proposed to measure the angular momentum about that axis.

\section{Conservation of angular momentum}
The main result of this paper is that
\begin{equation}
L_\xi J\cong-\oint_S{*}(T_{aB}+\Theta_{aB})\psi^a\tau^B \label{amc}
\end{equation}
holds along a trapping horizon under the conditions (\ref{div})--(\ref{lie}),
where
\begin{equation}
\Theta_{aB}=-\frac1{16\pi}h^{cd}D_d\sigma_{Bac} \label{Theta2}
\end{equation}
is thereby determined to be the transverse-normal block of the effective
energy tensor for gravitational radiation. It can be shown by differentiating
the angular momentum (\ref{am}) using (\ref{lie}) to give
\begin{equation}
L_\xi
J\cong\frac1{8\pi}\oint_S{*}(\theta_\xi\psi^a\omega_a+\psi^aL_\xi\omega_a)
\end{equation}
then expanding $\xi$ by (\ref{xi}) and using the twisting equations
(\ref{twisting}) to express $L_\xi\omega$. The term in
$\theta_\xi=\xi^+\theta_++\xi^-\theta_-$ cancels with the first term from
(\ref{twisting}), while the $D$ gradients may all be removed as total
divergences due to (\ref{dxi}), (\ref{div}) and the fact that (\ref{con})
reduces to $\psi^aD_a\theta_\mp\cong0$ on a trapping horizon with
$\theta_\pm\cong0$. This leaves just
\begin{eqnarray}
L_\xi
J&\cong&\frac1{8\pi}\oint_S{*}\psi^a\Big(\xi^+(\textstyle{\frac12}h^{cd}D_d\sigma_{+ac}-8\pi
T_{a+})\nonumber\\
&&\qquad\qquad\quad-\xi^-(\textstyle{\frac12}h^{cd}D_d\sigma_{-ac}-8\pi
T_{a-})\Big)
\end{eqnarray}
which is an expanded form of (\ref{amc}), noting (\ref{tau}) and thereby
identifying (\ref{Theta2}).

Apart from the inclusion of $\Theta$, the conservation law (\ref{amc}) is the
standard surface-integral form of conservation of angular momentum, were
$\psi$ an axial Killing vector. It thereby describes the increase or decrease
of angular momentum of a black hole due to infall of co-rotating or
counter-rotating matter respectively. The corresponding volume-integral form
for a spatial horizon $H$, expressing the change $[J]_{\partial H}$ in $J$
between two marginal surfaces, follows as
\begin{equation}
[J]_{\partial H}\cong-\int_H{\hat*}(T_{aB}+\Theta_{aB})\psi^a\hat\tau^B
\label{amcv}
\end{equation}
where $\hat\tau=\tau/\sqrt{g(\xi,\xi)}$ is the unit normal vector and
${\hat*}1 ={*}\sqrt{g(\xi,\xi)}\wedge dx$ is the proper volume element.
Although more familiar, as for the energy conservation law \cite{bhd2,bhd3},
this form becomes degenerate as the horizon becomes null, since ${\hat*}1\to0$
while $\hat\tau$ ceases to exist. Since this is a physically important limit,
where a black hole is not growing, the surface-integral form (\ref{amc}) is
preferred.

The null shears $\sigma_{\pm bc}$ have previously been identified in the
energy conservation law (\ref{ec}) as encoding the ingoing and outgoing
transverse gravitational radiation, via the energy densities (\ref{Theta0}),
which agree with expressions in other limits, such as the Bondi energy density
at null infinity and the Isaacson energy density of high-frequency linearized
gravitational waves \cite{bhd2,bhd3}. So the expression (\ref{Theta2}) implies
that gravitational radiation with a transversely differential waveform will
generally possess angular momentum density. One can see corresponding terms in
the linearized approximation \cite{MTW}, but they are set to zero by the
transverse-traceless gauge conditions, which are ``transverse'' in a different
sense to that used here. In any case, the conservation law shows that a black
hole can spin up or spin down even in vacuum, at a rate related to ingoing and
outgoing gravitational radiation.

The identification of the transverse-normal block (\ref{Theta2}) of $\Theta$
appears to be new. Previous versions of angular momentum flux laws for
dynamical black holes \cite{AK1,AK2,AK3,BF1,BF2,BF3,Gou} contain different
terms, which are not in energy-tensor form, i.e.\ some tensor contracted with
$\psi$ and $\tau$. They can be recovered by removing a transverse divergence
from $\Theta_{aB}\psi^a\tau^B$, yielding
$\sigma_\tau^{ad}D_d\psi_a/16\pi=\sigma_\tau^{ad}L_\psi h_{ad}/32\pi$, where
$\sigma_\tau^{ad}=\tau^Bh^{ae}h^{cd}\sigma_{Bce}$ encodes the shear along
$\tau$. Such terms have been described by analogy with viscosity
\cite{Gou,GJ2}.

The conservation laws (\ref{ec}) and (\ref{amc}) take a similar form,
expressing rate of change of mass $M$ and angular momentum $J$ as surface
integrals of densities of energy and angular momentum, with respect to
preferred vectors $k$ and $\psi$ which play the role of stationary and axial
Killing vectors, even if there are no symmetries. Of the ten conservation laws
in flat-space physics, they are the two independent laws expected for an
astrophysical black hole, which defines its own spin axis and centre-of-mass
frame, in which its momentum vanishes.

\section{Quasi-local conservation laws}
For an electromagnetic field with charge-current density vector $j$, the total
electric charge $Q$ in a region $H$ of a spatial hypersurface is defined as
\begin{equation}
[Q]_{\partial H}=-\int_H{\hat*}g(j,\hat\tau). \label{ccv}
\end{equation}
The surface-integral form of conservation of charge follows by the same
arguments relating (\ref{amc}) and (\ref{amcv}):
\begin{equation}
L_\xi Q=-\oint_S{*}g(j,\tau). \label{cc}
\end{equation}
As before, this is more general, since $H$ may have any signature. The
conservation laws for energy (\ref{ec}) and angular momentum (\ref{amc})
evidently take the same form
\begin{equation}
L_\xi M\cong-\oint_S{*}g(\tilde\jmath,\tau), \quad L_\xi
J\cong-\oint_S{*}g(\bar\jmath,\tau) \label{cl}
\end{equation}
by identifying current vectors
\begin{equation}
\tilde\jmath^B=-k_A(T^{AB}+\Theta^{AB}),
\quad\bar\jmath^B=\psi_a(T^{aB}+\Theta^{aB}).
\end{equation}

The local differential form of charge conservation,
\begin{equation}
\nabla\!_\alpha j^\alpha=0 \label{ccd}
\end{equation}
where $\nabla$ is the covariant derivative of $g$, notably does not hold for
$\tilde\jmath$ or $\bar\jmath$ in general. A weaker property holds, obtained
as follows. First note that in any of the three conservation laws (\ref{ec}),
(\ref{amc}) and (\ref{cc}), $\xi$ and $\tau$ may be interchanged. Thus there
are really two independent laws in each case. This can be understood from
special relativity: if $\xi$ were causal, one would interpret them as
expressing rate of change of energy, angular momentum or charge as,
respectively, power, torque or current; while if $\xi$ were spatial, one would
normally convert to volume-integral form and regard them as defining the
energy, angular momentum or charge in a region. One can make either
interpretation for a black hole, since $H$ would be generically spatial, but
the marginal surfaces $S$ locate the black hole in a family of time slices.

Given two independent equations in the normal space, it follows that
\begin{equation}
\epsilon^A{}_BL_AM\cong-\oint_S{*}1\wedge\tilde\jmath_B, \quad
\epsilon^A{}_BL_AJ\cong-\oint_S{*}1\wedge\bar\jmath_B.
\end{equation}
Expressed in terms of the curl and divergence of the normal space,
\begin{equation}
\hbox{curl}\,M\cong-\oint_S{*}1\wedge\tilde\jmath, \quad
\hbox{curl}\,J\cong-\oint_S{*}1\wedge\bar\jmath
\end{equation}
whereas
\begin{equation}
\nabla\!_\alpha\jmath^\alpha={*}\hbox{div}({*}1\wedge\jmath)
\end{equation}
for any normal vector $\jmath$. Then $\hbox{div}\,\hbox{curl}\,=0$ yields
\begin{equation}
\oint_S{*}\nabla\!_\alpha\tilde\jmath^\alpha
\cong\oint_S{*}\nabla\!_\alpha\bar\jmath^\alpha\cong0. \label{qlc}
\end{equation}
This subtly confirms the view that energy and angular momentum in General
Relativity cannot be localized \cite{MTW}, but might be quasi-localized, as
surface integrals \cite{Pen}. The corresponding conservation laws have indeed
been obtained in surface-integral but not local form.

\section{State space}
There are now three conserved quantities $(M,J,Q)$, forming a state space for
dynamical black holes. Following various authors
\cite{ABL1,Boo,AK1,AK2,AK3,BF1}, related quantities may then be defined by
formulas satisfied by Kerr-Newman black holes, specifically those for the ADM
energy
\begin{equation} E\cong\frac{\sqrt{((2M)^2+Q^2)^2+(2J)^2}}{4M} \label{e}
\end{equation}
the surface gravity
\begin{equation}
\kappa\cong\frac{(2M)^4-(2J)^2-Q^4}{2(2M)^3\sqrt{((2M)^2+Q^2)^2+(2J)^2}}
\label{sg}
\end{equation}
the angular speed
\begin{equation}
\Omega\cong\frac{J}{M\sqrt{((2M)^2+Q^2)^2+(2J)^2}} \label{as}
\end{equation}
and the electric potential
\begin{equation}
\Phi\cong\frac{((2M)^2+Q^2)Q}{2M\sqrt{((2M)^2+Q^2)^2+(2J)^2}}. \label{ep}
\end{equation}
It would be preferable to have independently motivated definitions of these
quantities, but so far this has been done only in spherical symmetry
\cite{sph,1st,in}, where there are natural definitions of $E$, $\kappa$ and
$\Phi=Q/R$ which can be applied anywhere in the space-time, coinciding with
the above expressions on the outer horizons of a Reissner-Nordstr\"om black
hole.

In the dynamical context, $E\ge M$ is generally not the ADM energy, since
there may be matter or gravitational radiation outside the black hole. Rather,
it can be interpreted as the effective energy of the black hole, as follows.
Defining the moment of inertia $I$ by the usual formula $J\cong I\Omega$ yields
\begin{equation}
I\cong M\sqrt{((2M)^2+Q^2)^2+(2J)^2}\cong ER^2.
\end{equation}
Expanding $E$ for $J\ll M^2$ and $Q\ll M$ yields
\begin{equation}
E\approx M+\textstyle{\frac12}I\Omega^2+\textstyle{\frac12}Q^2/R.
\end{equation}
The second and third terms are standard expressions for rotational kinetic
energy and electrostatic energy. Thus the irreducible mass $M$ plays the role
of a rest mass, with $E$ including contributions from rotational and
electrical energy.

Returning to the general case, the above definitions satisfy the state-space
formulas
\begin{equation}
\kappa\cong8\pi\frac{\partial E}{\partial A}\cong{1\over{4M}}\frac{\partial
E}{\partial M}, \quad\Omega\cong\frac{\partial E}{\partial J},
\quad\Phi\cong\frac{\partial E}{\partial Q}. \label{pd}
\end{equation}
There follows a dynamic version of the so-called first law of black-hole
mechanics \cite{BCH}:
\begin{equation}
L_\xi E\cong\frac{\kappa}{8\pi}L_\xi A+\Omega L_\xi J+\Phi L_\xi Q.
\end{equation}
As desired, the state-space perturbations in the classical law for Killing
horizons \cite{BCH}, or the version for isolated horizons \cite{ABD,ABL1,Boo},
have been replaced by the derivatives along the trapping horizon, thereby
promoting it to a genuine dynamical law.

The rate of change of effective energy can also be written in energy-tensor
form,
\begin{equation}
L_\xi
E=\oint_S{*}\left((T_{\alpha\beta}+\Theta_{\alpha\beta})K^\alpha\tau^\beta-\Phi
j_\beta\tau^\beta\right)
\end{equation}
where
\begin{equation}
K=4M\kappa k-\Omega\psi
\end{equation}
plays the role of the stationary Killing vector. Note that in the classical
theory of stationary black holes, the state variables are usually taken to be
$(E,J,Q)$, with the irreducible mass $M$ defined as a dependent variable. The
dynamical theory reveals that $(M,J,Q)$ are the more basic variables, since
they each satisfy a simple conservation law. Then the effective energy $E$ is
defined as a dependent variable and therefore satisfies the above conservation
law. This reflects a shift in emphasis from the classical to the dynamical
theory: the so-called first law is a dependent result, obtained from more
fundamental conservation laws for energy, angular momentum and charge.

\section{Null trapping horizons and zeroth law}
A zeroth law for trapping horizons follows from the above, if one defines
local equilibrium by the absence of relevant fluxes:
\begin{equation}
g(\tilde\jmath,\tau)\cong g(\bar\jmath,\tau)\cong g(j,\tau)\cong0.\label{0th}
\end{equation}
Then $(M,J,Q)$ are constant on the horizon and so is $\kappa$. In fact, these
conditions do hold on a null trapping horizon under the dominant energy
condition, as shown below. This treatment also turns out to be compatible with
the definition of weakly isolated horizon \cite{AFK,ABL1,Boo,ABL2,DKSS,GJ1}.

Consider a null trapping horizon, assumed henceforth in this section to be
given by $\theta_+\cong0$. The null focussing equation yields
$T_{++}+||\sigma_+||^2/32\pi\cong0$, so the null energy condition, which
implies $T_{++}\ge0$, yields \cite{bhd}
\begin{equation}
T_{++}\cong0,\quad\sigma_+\cong0.\label{ed}
\end{equation}
Thus the degenerate metric of the horizon is preserved along the generating
vector. The dominant energy condition, which implies that the energy-momentum
$P_\alpha=-T_{\alpha\beta}l_+^\beta$ is causal, further yields \cite{AFK}
\begin{equation}
T_{+a}\cong0\label{amd}
\end{equation}
since $P=-T_{++}dx^+-T_{+-}dx^--T_{+a}dx^a$ would otherwise be spatial.

On a null hypersurface $H$, one can take the null coordinate to be the
generating coordinate, $x^+\cong x$, meaning that the shift vector vanishes,
$s_+\cong0$, so that $\xi\cong\tau\cong l_+$. Since $k\cong-e^fL_-Rl_+$
(\ref{time}), one finds
\begin{equation}
g(\tilde\jmath,\tau)\cong e^fL_-R(T_{++}+\Theta_{++}),\quad
g(\bar\jmath,\tau)\cong\psi^a(T_{a+}+\Theta_{a+}).
\end{equation}
These fluxes vanish by the above results (\ref{ed})--(\ref{amd}) and the
expressions (\ref{Theta0}), (\ref{Theta2}) for components of $\Theta$. The
other flux in (\ref{0th}) vanishes due to the Maxwell equations \cite{AFK}.

For a null trapping horizon, the dual-null foliation is not unique, so the
question arises whether there is a natural way to fix it. An affirmative
answer is given by noting that the above results also imply, via the twisting
equations (\ref{twisting}),
\begin{equation}
L_+(\omega-{\textstyle\frac12}Df)\cong0.
\end{equation}
This restriction on the dual-null geometry is suggestive of a
proto-conservation law for angular momentum. Now the only normal fundamental
form intrinsic to a null hypersurface is $\zeta_{-+}=\frac12Df+\omega$ of
Appendix B, which was therefore used by Ashtekar et al.\ \cite{AFK} to define
angular momentum for an isolated horizon. However, it is the other null normal
fundamental form $\zeta_{+-}=\frac12Df-\omega$, depending on the dual-null
foliation, which is preserved as above. Given that the general definition
(\ref{am}) of angular momentum involves $\omega$, it seems best to fix the
unwanted freedom by
\begin{equation}
Df\cong0.\label{fix}
\end{equation}
Recalling the definition (\ref{ff}) or (\ref{ls}) of $f$, this condition fixes
the normalization of the extrinsic null normal $l_-$ with respect to the
intrinsic null normal $l_+$, which is always possible on a null hypersurface
$H$. In fact, it is common simply to fix $f\cong0$. A similar normalization is
also used in the context of null infinity.

Then the definition (\ref{am}) of angular momentum becomes unambiguous on a
null trapping horizon, coincides with the definition for isolated horizons
\cite{ABD,ABL1,Boo}, and is preserved along the horizon, assuming only that
$\psi$ is a coordinate vector field (\ref{lie}):
\begin{equation}
L_\xi J\cong0.
\end{equation}
Since the area law \cite{bhd} shows that $A$ is increasing unless $H$ is null
everywhere on a given $S$, this answers, in the negative, a simply stated
physical question: whether a black hole can change its angular momentum
without increasing its area.

The above reasoning has largely recovered the notion of a weakly isolated
horizon introduced by Ashtekar et al.\ \cite{AFK}, except that the scaling
freedom in $l_+$ has not been fixed. In more detail, Ashtekar et al.\
\cite{AFK} introduced a 1-form which will here be denoted by $\varpi$, defined
by
\begin{equation}
\hat\nabla_\alpha l_+^\beta=\varpi_\alpha l_+^\beta
\end{equation}
where $\hat\nabla$ is the covariant derivative operator of $H$. The transverse
and normal components are found as
\begin{equation}
\bot\varpi=-\omega-{\textstyle\frac12}Df,\quad l_+^\alpha\varpi_\alpha=-\nu_+.
\end{equation}
Since $\varpi$ is an invariant of $H$ and $l_+$, Ashtekar et al.\ \cite{AFK}
demanded
\begin{equation}
L_+\varpi\cong0
\end{equation}
in order to define a weakly isolated horizon. The transverse part agrees with
the above results, which also imply $D\nu_+\cong0$, while the normal part
further fixes the inaffinity $\nu_+$ to be constant on $H$. This fixes the
scaling of $l_+$ up to a constant multiple. Ashtekar et al.\ \cite{AFK}
defined the surface gravity to be
\begin{equation}
\hat\kappa\cong-\nu_+
\end{equation}
which recovers the standard surface gravity of a Killing horizon if $l_+$ is
the null Killing vector \cite{Wal}. Then the constancy of $\hat\kappa$ can
also be interpreted as a zeroth law. This still leaves non-zero $\hat\kappa$
ambiguous up to a constant multiple, not necessarily agreeing with the
definition (\ref{sg}), which therefore fixes that freedom.

The above considerations appear to have a closed a gap in the general
paradigm, concerning how a growing black hole ceases to grow. It seems that
the generically spatial trapping horizon simply becomes null. It is difficult
to find a practical formalism describing all cases without some degeneracy
arising in the null case, but the dual-null formalism appears to be adequate;
one fixes the additional freedom in the null case by (\ref{fix}). In
particular, no additional conditions need be imposed on the horizon itself, as
compared with the variety of definitions of isolated horizons
\cite{ABF1,ABF2,AFK,ABD,ABL1,Boo,ABL2,DKSS,GJ1}. Numerical evidence that such
horizons exist in practice has been given by Dreyer et al.\ \cite{DKSS}, who
looked for and found approximately null trapping horizons, under the name
non-expanding horizons.

Research supported by the National Natural Science Foundation of China under
grants 10375081 and 10473007 and by Shanghai Normal University under grant
PL609. Thanks to Abhay Ashtekar, Ivan Booth, Eric Gourgoulhon and Badri
Krishnan for discussions.

\appendix
\section{Twist and weak fields}
The twist may be calculated by first finding the shift vectors $s_\pm$, due to
the form \cite{dne}
\begin{equation}
\omega={\textstyle\frac12}e^fh\left([\partial_+,s_-]-[\partial_-,s_+]+[s_-,s_+]\right).
\end{equation}
If it is more convenient to use an orthonormal basis $\{l_0,l_1\}$ of the
normal space,
\begin{equation}
\bot l_0=\bot l_1=0=g(l_0,l_1),\quad g(l_0,l_0)=-1=-g(l_1,l_1)\label{ob}
\end{equation}
then
\begin{equation}
\omega={\textstyle\frac12}h\left([l_0,l_1]\right)
\end{equation}
follows by linear combinations from (\ref{am}), or directly from the Komar
integral (\ref{komar}). If the basis is adapted to a coordinate basis via
\begin{equation}
l_A=\partial_A-s_A,\quad s_A=\bot\partial_A
\end{equation}
then
\begin{equation}
\omega={\textstyle\frac12}h\left([\partial_1,s_0]-[\partial_0,s_1]+[s_0,s_1]\right).
\end{equation}
In either case, the first step is to find the shift vectors.

The weak-field metric \cite{MTW}, in standard spherical polar coordinates
$(t,r,\vartheta,\varphi)$ adapted to the axis of rotation, is
\begin{eqnarray}
g&\sim&-\left(1-\frac{2M}r\right)dt^2-\frac{4J}r\sin^2\vartheta
dtd\varphi\nonumber\\
&&+\left(1+\frac{2M}r\right)\left(dr^2+r^2(d\vartheta^2+\sin^2\vartheta
d\varphi^2)\right)
\end{eqnarray}
where, in this appendix only, $M$ and $J$ denote the mass and angular momentum
as defined in this approximation, obtained by linearizing the metric and
neglecting higher powers of $1/r$. The inverse metric is
\begin{eqnarray}
g^{-1}&\sim&-\left(1+\frac{2M}r\right)\partial_t^2-\frac{4J}{r^3}\partial_t\partial_\varphi\nonumber\\
&&+\left(1-\frac{2M}r\right)\left(\partial_r^2+\frac1{r^2}
\left(\partial_\vartheta^2+\frac{\partial_\varphi^2}{\sin^2\vartheta}\right)\right).
\end{eqnarray}
Taking the transverse surfaces $S$ as those of constant $(t,r)$, one can read
off the nonzero component of the shift 1-forms $s_{Ab}$ as
\begin{equation}
s_{t\varphi}=g_{t\varphi}\sim-\frac{2J}r\sin^2\vartheta,\quad
s_t{}^\varphi\sim-\frac{2J}{r^3}.
\end{equation}
Then the non-zero component of the twist is given by
\begin{equation}
\omega^\varphi\sim{\textstyle\frac12}\partial_rs_t{}^\varphi\sim\frac{3J}{r^4},
\quad\omega_\varphi\sim\frac{3J}{r^2}\sin^2\vartheta.
\end{equation}
The area form is
\begin{equation}
{*}1\sim r^2\sin\vartheta d\vartheta\wedge d\varphi
\end{equation}
so that
\begin{equation}
{*}\omega_\varphi\sim3J\sin^3\vartheta d\vartheta\wedge d\varphi.
\end{equation}
Standard integrals yield
\begin{equation}
\oint_S{*}\omega_\varphi\sim8\pi J.
\end{equation}
Since $\omega_a\psi^a=\omega_\varphi$ if $\psi=\partial/\partial\varphi$, this
agrees with the general definition (\ref{am}) of angular momentum.

A directly measurable quantity due to rotational frame-dragging is the
precessional angular velocity \cite{MTW}
\begin{equation}
\vec\Omega_{LT}\sim\frac{J}{r^3}\left(3(\hat z\cdot\hat r)\hat r-\hat z\right)
\end{equation}
of a gyroscope due to the Lense-Thirring effect, where $\hat r$ is a unit
vector in the direction of the gyroscope and $\hat z$ is a unit vector along
the axis of rotation. Results of measurements of the effect due to the Earth
by Gravity Probe B are expected soon. If the twist $\omega$ is formally
converted to an angular velocity $\vec\Omega$ by
\begin{equation}
\omega\sim\vec\Omega\times\hat r
\end{equation}
then
\begin{equation}
\vec\Omega\sim\left|\frac\omega{\sin\vartheta}\right|\hat z
\sim\frac{3J}{r^3}\hat z
\end{equation}
does have the direction and relativistic dimensions of angular velocity. Then
\begin{equation}
\vec\Omega_{LT}\sim(\vec\Omega\cdot\hat r)\hat r-{\textstyle\frac13}\vec\Omega.
\end{equation}
Curiously, this is a linear transformation of $\vec\Omega$, by the same
traceless tensor used in defining quadrupole moments \cite{MTW}. The puckish
role of the factor of 3 in the above calculations is also noteworthy. In any
case, it confirms that the twist does indeed encode the twisting around of
space-time due to a rotating mass, in a directly measurable way.

For completeness, the agreement of the weak-field mass with the Hawking mass
can also be checked, as follows. One needs to keep track of an extra power of
$1/r$ in the area form
\begin{equation}
{*}1\sim\left(1+\frac{2M}r\right)r^2\sin\vartheta d\vartheta\wedge d\varphi.
\end{equation}
Then the expansion 1-form $\theta_A$ has non-zero component
\begin{equation}
\theta_r\sim\partial_r\log\left(\left(1+\frac{2M}r\right)r^2\right)
\sim\frac2r\left(1-\frac Mr\right).
\end{equation}
Then
\begin{equation}
\gamma^{AB}\theta_A\theta_B\sim\gamma^{rr}\theta_r\theta_r
\sim\frac4{r^2}\left(1-\frac{4M}r\right)
\end{equation}
and
\begin{eqnarray}
\oint_S{*}\gamma^{AB}\theta_A\theta_B&\sim&
\oint_S4\left(1-\frac{2M}r\right)\sin\vartheta d\vartheta\wedge d\varphi\nonumber\\
&\sim&16\pi\left(1-\frac{2M}r\right).
\end{eqnarray}
Since $R\sim r$, this agrees with the Hawking mass (\ref{hm}).

\section{Normal fundamental forms}
Various definitions of angular momentum \cite{AK1,ABD,BY} are similar to
(\ref{am}), with $\omega$ replaced by a 1-form which is, implicitly or
explicitly, a normal fundamental form. To clarify the situation, normal
fundamental forms are reviewed below, referring to previous treatments
\cite{dne,Gou}.

Writing the twist (\ref{tw}) explicitly in components,
\begin{equation}
2\omega_\alpha=e^fh_{\alpha\beta}(l_-^\gamma\nabla_\gamma
l_+^\beta-l_+^\gamma\nabla_\gamma l_-^\beta).
\end{equation}
If $l_\pm$ are adapted to a coordinate basis via (\ref{nn}), then
commutativity $\nabla_{[\beta}\nabla_{\gamma]}x^\pm=0$ allows it to be written
as
\begin{equation}
2\omega_\alpha=e^fh_\alpha^\beta(l_-^\gamma\nabla_\beta
l_{+\gamma}-l_+^\gamma\nabla_\beta l_{-\gamma}).
\end{equation}
This is the difference of two normal fundamental forms $\zeta_{\mp\pm}$ with
components
\begin{equation}
\zeta_{\mp\pm\alpha}=e^fl_\mp^\gamma h_\alpha^\beta\nabla_\beta l_{\pm\gamma}.
\end{equation}
They are the independent normal fundamental forms, since the corresponding
$\zeta_{\pm\pm}$ vanish. Their sum is $Df$, since the normalization in
(\ref{ls}) yields
\begin{equation}
D_\alpha f=e^fh_\alpha^\beta(l_-^\gamma\nabla_\beta
l_{+\gamma}+l_+^\gamma\nabla_\beta l_{-\gamma}).
\end{equation}
Then
\begin{equation}
\zeta_{\mp\pm}={\textstyle\frac12}Df\pm\omega
\end{equation}
and the normal fundamental forms are thereby encoded in $\omega$ and $Df$.

For an orthonormal basis (\ref{ob}), there is just one independent normal
fundamental form $\hat\zeta_{01}=-\hat\zeta_{10}$, given by
\begin{equation}
\hat\zeta_{01\alpha}=l_0^\gamma h_\alpha^\beta\nabla_\beta l_{1\gamma},
\quad\hat\zeta_{10\alpha}=l_1^\gamma h_\alpha^\beta\nabla_\beta l_{0\gamma}
\end{equation}
with corresponding $\hat\zeta_{00}$, $\hat\zeta_{11}$ vanishing. Under a boost
transformation
\begin{equation}
l_0\mapsto l_0\cosh\rho+l_1\sinh\rho,\quad l_1\mapsto l_0\sinh\rho+l_1\cosh\rho
\end{equation}
which preserves the orthonormal conditions (\ref{ob}), the normal fundamental
form is generally not invariant:
\begin{equation}
\hat\zeta_{01}\mapsto\hat\zeta_{10}-D\rho,\quad\hat\zeta_{10}\mapsto\hat\zeta_{10}+D\rho.
\end{equation}
However, if the orthonormal basis is adapted to the dual-null basis by, e.g.\
$\sqrt2l_\pm=l_0+l_1$, then
\begin{equation}
\hat\zeta_{01}=-\hat\zeta_{10}=\omega.
\end{equation}
For spatial $H$, the same result is obtained by adapting the orthonormal basis
by choosing
\begin{equation}
l_0=\tau/\sqrt{g(\xi,\xi)},\quad l_1=\xi/\sqrt{g(\xi,\xi)}.\label{aon}
\end{equation}
The missing information in $Df$ can be recovered by instead defining
\begin{eqnarray}
\zeta_{01\alpha}&=&\frac{\tau^\gamma
h_\alpha^\beta\nabla_\beta\xi_\gamma}{g(\xi,\xi)},
\quad\zeta_{10\alpha}=\frac{\xi^\gamma
h_\alpha^\beta\nabla_\beta\tau_\gamma}{g(\xi,\xi)},\nonumber\\
\zeta_{00\alpha}&=&\frac{\tau^\gamma
h_\alpha^\beta\nabla_\beta\tau_\gamma}{g(\xi,\xi)},
\quad\zeta_{11\alpha}=\frac{\xi^\gamma
h_\alpha^\beta\nabla_\beta\xi_\gamma}{g(\xi,\xi)}.
\end{eqnarray}
Then
\begin{equation}
\zeta_{01}=-\zeta_{10}=\omega,\quad\zeta_{00}=-\zeta_{11}={\textstyle\frac12}Df.
\end{equation}
These 1-forms nevertheless become degenerate when $\xi$ becomes null.

As shown explicitly by Gourgoulhon \cite{Gou}, the 1-form used by Brown \&
York \cite{BY} to define angular momentum is, in this notation,
$-\hat\zeta_{10}$. Therefore it coincides with $\omega$ if the orthonormal
basis is adapted to a trapping horizon, but generally not if it is adapted to
a foliation of spatial hypersurfaces intersecting the trapping horizon in the
marginal surfaces. The 1-form used by Ashtekar \& Krishnan \cite{AK1} to
define angular momentum for dynamical horizons is also $-\hat\zeta_{10}$, this
time explicitly adapted to the horizon by (\ref{aon}), therefore coinciding
with $\omega$ in this case. The 1-form used by Ashtekar et al.\
\cite{ABD,ABL1,Boo} to define angular momentum for isolated horizons is
$\zeta_{-+}$, which generally does not coincide with $\omega$. However, it does
coincide if the gauge freedom is fixed by (\ref{fix}).

\section{Kerr example}
Consider a Kerr space-time in Boyer-Lindquist coordinates
$(t,r,\vartheta,\varphi)$, with $S$ given by constant $(t,r)$ and
$\xi=\partial/\partial r$. Then
\begin{equation}
{*}1=\sqrt{(r^2+a^2)^2-\Delta a^2\sin^2\vartheta}\sin\vartheta
d\vartheta\wedge d\varphi
\end{equation}
where $\Delta=r^2-2mr+a^2$,
\begin{equation}
\theta_\xi=\frac{2r(r^2+a^2)-(r-m)a^2\sin^2\vartheta}{(r^2+a^2)^2-\Delta
a^2\sin^2\vartheta}
\end{equation}
and
\begin{equation}
D\theta_\xi=\frac{2a^2(r^2+a^2)(r^3-3mr^2+a^2r+ma^2)\sin\vartheta\cos\vartheta}{((r^2+a^2)^2-\Delta
a^2\sin^2\vartheta)^2}d\vartheta.
\end{equation}
If $a\not=0$, $D\theta_\xi$ is non-zero except at the poles, equator and
isolated values of $r$, so the construction yields a unique continuous $\psi$,
$\psi=\partial/\partial\varphi$.


\begin{thebibliography}{99}
\bibitem{Sch}K Schwarzschild,
 {Sitzber. Deut. Akad. Wiss. Berlin, Kl. Math.-Phys. Tech.} 189 (1916).
\bibitem{Ein}A Einstein, {Preuss. Akad. Wiss. Berlin, Sitzber.} 844 (1915).
\bibitem{Rei}H Reissner, {Ann. Physik} {\bf 50}, 106 (1916).
\bibitem{Nor}G Nordstr\"om, {Proc. Kon. Ned. Akad. Wet.} {\bf 20}, 1238 (1918).
\bibitem{ER}A Einstein \& N Rosen, {Phys. Rev.} {\bf 48}, 73 (1935).
\bibitem{OS}J R Oppenheimer \& H Snyder, {Phys. Rev.} {\bf 56}, 455 (1939).
\bibitem{Kru}M D Kruskal, {Phys. Rev.} {\bf 119}, 1743 (1960).
\bibitem{WF}J A Wheeler \& K Ford, Geons, Black Holes \& Quantum Foam:
 A Life in Physics (W W Norton 2000).
\bibitem{Ker}R P Kerr, {Phys. Rev. Lett.} {\bf 11}, 237 (1963).
\bibitem{New}E T Newman, E Couch, K Chinnapared, A Exton, A Prakash \& R
Torrence, {J. Math. Phys.} {\bf 6}, 918 (1965).
\bibitem{Whe}J A Wheeler, {Amer. Sci.} {\bf 56}, 1 (1968).
\bibitem{Pen0}R Penrose, in Battelle Rencontres
 ed.\ C M DeWitt \& J A Wheeler (Benjamin 1968).
\bibitem{Haw0}S W Hawking, {Phys. Rev. Lett.} {\bf 26}, 1344 (1971).
\bibitem{BCH}J M Bardeen, B Carter \& S W Hawking,
 {Comm. Math. Phys.} {\bf 31}, 161 (1973).
\bibitem{HE}S W Hawking \& G F R Ellis,
 The Large Scale Structure of Space-Time (Cambridge University Press 1973).
\bibitem{MTW}C W Misner, K S Thorne \& J A Wheeler,
 Gravitation (Freeman 1973).
\bibitem{Wal}R M Wald, General Relativity (University of Chicago Press 1984).
\bibitem{bhd}S A Hayward, {Phys. Rev.} {\bf D49}, 6467 (1994).
\bibitem{sph}S A Hayward, {Phys. Rev.} {\bf D53}, 1938 (1996).
\bibitem{1st}S A Hayward, {Class. Quantum Grav.} {\bf 15}, 3147 (1998).
\bibitem{in}S A Hayward, {Phys. Rev. Lett.} {\bf 81}, 4557 (1998).
\bibitem{cyl}S A Hayward, {Class. Quantum Grav.} {\bf 17}, 1749 (2000).
\bibitem{qs}S A Hayward, {Phys. Rev.} {\bf D61}, 101503 (2000).
\bibitem{SH}H Shinkai \& S A Hayward,
 {Phys. Rev.} {\bf D64}, 044002 (2001).
\bibitem{gwbh}S A Hayward, {Class. Quantum Grav.} {\bf 18}, 5561 (2001).
\bibitem{gwe}S A Hayward, {Phys. Lett.} {\bf A294}, 179 (2002).
\bibitem{ABF1}A Ashtekar, C Beetle \& S Fairhurst,
 {Class. Quantum Grav.} {\bf 16}, L1 (1999).
\bibitem{ABF2}A Ashtekar, C Beetle \& S Fairhurst,
 {Class. Quantum Grav.} {\bf 17}, 253 (2000).
\bibitem{AFK}A Ashtekar, S Fairhurst \& B Krishnan,
 {Phys. Rev.} {\bf D62}, 104025 (2000).
\bibitem{ABD}A Ashtekar, C Beetle, O Dreyer, S Fairhurst, B Krishnan, J
Lewandowski \& J Wisniewski,
 {Phys. Rev. Lett.} {\bf 85}, 3564 (2000).
\bibitem{ABL1}A Ashtekar, C Beetle \& J Lewandowski,
 {Phys. Rev.} {\bf D64}, 044016 (2001).
\bibitem{Boo}I Booth, {Class. Quantum Grav.} {\bf 18}, 4239 (2001).
\bibitem{ABL2}A Ashtekar, C Beetle \& J Lewandowski,
 {Class. Quantum Grav.} {\bf 19}, 1195 (2002).
\bibitem{DKSS}O Dreyer, B Krishnan, D Shoemaker \& E Schnetter,
 {Phys. Rev.} {\bf D67}, 024018 (2003).
\bibitem{GJ1}E Gourgoulhon \& J L Jaramillo, {Phys. Rep.} {\bf 423}, 159 (2006).
\bibitem{AK1}A Ashtekar \& B Krishnan,
 {Phys. Rev. Lett.} {\bf 89}, 261101 (2002).
\bibitem{AK2}A Ashtekar \& B Krishnan, {Phys. Rev.} {\bf D68}, 104030 (2003).
\bibitem{AK3}A Ashtekar \& B Krishnan, {Living Rev. Relativity} {\bf 7}, 10
(2004).
\bibitem{BF1}I Booth \& S Fairhurst, {Phys. Rev. Lett.} {\bf 92}, 011102 (2004).
\bibitem{BF2}I Booth \& S Fairhurst, {Class. Quantum Grav.} {\bf 22}, 4515 (2005).
\bibitem{BF3}I Booth \& S Fairhurst, Isolated, slowly evolving and dynamical trapping horizons:
geometry and mechanics from surface deformations, gr-qc/0610032.
\bibitem{AMS}L Andersson, M Mars \& W Simon,  {Phys. Rev. Lett.} {\bf 95}, 111102 (2005).
\bibitem{AG}A Ashtekar \& G J Galloway, {Adv. Theor. Math. Phys.} {\bf 9}, 1
 (2005).
\bibitem{dn}S A Hayward, {Ann. Inst. H. Poincar\'e} {\bf 59}, 399 (1993).
\bibitem{dne}S A Hayward, {Class. Quantum Grav.} {\bf 10}, 779 (1993).
\bibitem{bhd2}S A Hayward, {Phys. Rev. Lett.} {\bf 93}, 251101 (2004).
\bibitem{bhd3}S A Hayward, {Phys. Rev.} {\bf D70}, 104027 (2004).
\bibitem{Haw}S W Hawking, {J. Math. Phys.} {\bf 9}, 598 (1968).
\bibitem{j15}S A Hayward,
 Angular momentum and conservation laws for dynamical black holes,
 gr-qc/0601058.
\bibitem{bhd4}S A Hayward, Conservation laws for dynamical black holes
 gr-qc/0607081.
\bibitem{Kom}A Komar, {Phys. Rev.} {\bf 113}, 934 (1959).
\bibitem{BHMS}H Bray, S A Hayward, M Mars \& W Simon,
 Generalized inverse mean curvature flows in space-time, gr-qc/0603014.
\bibitem{gr}S A Hayward, {Class. Quantum Grav.} {\bf 23}, L15 (2006).
\bibitem{BY}J D Brown \& J W York, {Phys. Rev.} {\bf D47}, 1407 (1993).
\bibitem{Gou}E Gourgoulhon, {Phys. Rev.} {\bf D72}, 104007 (2005).
\bibitem{GJ2}E Gourgoulhon \& J L Jaramillo,
 Area evolution, bulk viscosity and entropy principles for dynamical horizons,
 gr-qc/0607050.
\bibitem{Pen}R Penrose, {Proc. R. Soc. London} {\bf A381}, 53 (1982).
\end{thebibliography}
\end{document}